\documentclass[a4paper]{article}

\usepackage{INTERSPEECH2015}

\usepackage{graphicx}
\usepackage{amssymb,amsmath,bm}
\usepackage{textcomp}
\usepackage{subfigure}

 \usepackage{algorithm}
\usepackage{algpseudocode}
\usepackage{stackengine}
\usepackage{enumitem}
\ninept

\title{A dictionary learning and source recovery based approach to classify diverse audio sources}

\makeatletter
\def\name#1{\gdef\@name{#1\\}}
\makeatother \name{{\em K V Vijay Girish$^1$,   T V Ananthapadmanabha$^2$ and A G Ramakrishnan$^1$, \textit{Senior Member,~IEEE.} }}

\address{$^1$ Department of Electrical Engineering,  Indian Institute of Science, Bangalore \\
 $^2$Voice and Speech Systems, Malleswaram, Bangalore, India \\ 
  {\small \tt kv@ee.iisc.ernet.in, tva.blr@gmail.com, ramkiag@ee.iisc.ernet.in}
}

\begin{document}
\maketitle
  \begin{abstract}
    A dictionary learning based audio source classification algorithm is proposed
    to classify a sample audio signal as one amongst a finite set of different
    audio sources. Cosine
    similarity measure is used to select the atoms during dictionary learning.
    Based on three objective measures proposed, namely, signal to distortion ratio (SDR),
    the number of non-zero weights and the sum of weights, a frame-wise source
    classification accuracy
    of 98.2\% is obtained for twelve different sources. Cent percent accuracy has
    been obtained using moving SDR accumulated over six
    successive frames for ten of the audio sources tested, while the two other
    sources require accumulation of 10 and 14 frames.
  \end{abstract}

\section{Introduction}
\subsection{Motivation for the present study}
 In techniques for speech enhancement, many a times the noise is assumed to be stationary with a known distribution. However, in a real world scenario, the noise may be non-stationary or speech may be corrupted  with different kinds of noises. The nature of noise varies with the environment such as traffic, restaurant, railway and bus station. Even competing speakers and music may impair intelligibility of speech. In the case of speech enhancement \cite{hu} and noise source separation, especially for hearing impaired \cite{baby,turner} the suppression of background audio for improving the intelligibility of speech would be more effective, if the type of  audio source can be classified. Other interesting application areas are forensics \cite{ikram}, machinery noise diagnostics \cite{lyon}, robotic navigation systems \cite{chu1} and acoustic signature classification  of aircrafts or vehicles \cite{shirkhodaie}.

This paper addresses the basic problem of classification of the type of audio from a finite set of sources, mostly noise and a couple of musical instruments. Noise classification can be seen as a first step in machine listening \cite{malkin}, which enables the system to know the background environment. Classification of noise types has been reported in the case of pure noise sources. Kates \cite{kates} addressed the problem of noise classification for hearing aid applications based on variation of signal envelope as features. Maleh et al. \cite{maleh} used line spectral frequencies as features for  classification of different kinds of noise as well as noise and speech classification. Casey \cite{casey} proposed a system to classify twenty different types of sounds using a hidden Markov model classifier and a reduced-dimension log-spectral features. Chu et al. \cite{chu} recognized 14 different environmental sounds using matching pursuit based features combined with mel-frequency cepstrum coefficients. Giannoulis et al. \cite{giannoulis} conducted a public evaluation challenge on acoustic scene classification (similar to noise classification), where 11 algorithms were evaluated along with a baseline system. The algorithms use time and frequency domain features extracted from the audio signal followed by a statistical model based  or majority vote based classifier. Cauchi \cite{cauchi} used non-negative matrix factorization for classification of auditory scenes.
  
      Representation of audio signals as a linear combination of non-negative sparse vectors called as dictionary atoms has been used for audio source separation \cite{virtanen0,ozerov,mysore}, recognition \cite{bertin,gemmeke,raj}, classification \cite{cho,zubair} and coding \cite{nikunen,plumbley}.   In this work, we only address the problem of  audio classification of pure noise sources using sparse non-negative representation of audio by proposing a novel dictionary learning  and a source recovery method. However, audio classification also works in a mixed audio signal, where segments have higher noise energy than speech.
     
\subsection{Review of Dictionary learning and source recovery}
    A dictionary is a matrix $D\in \rm I\!R^{p\times K}$ (with $p$ as the dimension of the acoustic feature vector) containing $K$ column vectors called atoms, denoted as $\mathbf{d_n}, 1\leq n\leq K$. A feature for any real valued  signal can be represented as $\mathbf{y}\approx D\mathbf{x}$, where $\mathbf{x}\in \rm I\!R^K$ is the vector containing weights for each dictionary atom. The vector $\mathbf{x}$ is estimated by minimizing the distance $dist(\mathbf{y},D\mathbf{x})$, where $dist()$ is a distance metric between $\mathbf{y}$ and $D\mathbf{x}$ such as $L_p$ norm or Kullback-Leibler (KL)-divergence \cite{virtanen}. In case the dictionary $D$ is overcomplete, the weight vector $x$ tends to be sparse. This method of estimating weights is termed as  sparse coding or source recovery. Matching pursuit \cite{mallat}, orthogonal matching pursuit (OMP) \cite{pati}, basis pursuit \cite{chen} and focal underdetermined system solver (FOCUSS) \cite{gorodnitsky} are some of the source recovery algorithms. 
    
    The simplest dictionary learning (DL) method is a random selection of observations from the training data \cite{virtanen}. K-means clustering \cite{coates} has also been used for DL. The relation between vector quantization and DL was shown by \cite{delgado}. Initial work on DL was carried out by Olshausen \cite{olshausen} and Lewicki \cite{lewicki} using probabilistic model of the features. Engal et al. \cite{engan} performed DL using a simple dictionary update (minimization of mean square error of the error matrix) and sparse coding using OMP or FOCUSS. Recursive least squares dictionary learning (RLS-DLA) \cite{skretting}, K-SVD \cite{aharon}, simultaneous codeword optimisation (SimCO) \cite{dai} and fast dictionary learning \cite{jafari} are other algorithms. DL and source recovery methods have been used for classification of objects in images by learning class-specific dictionaries \cite{kong}. Shafiee et al. \cite{shafiee} have used three different DL methods to classify faces and digits in images.

     In our work, we have adopted the recently reported active-set Newton algorithm (ASNA)\cite{virtanen} for source recovery. The training phase for the audio classification problem is DL from various noise/instrument sources. The advantage of this approach is that the audio sources need not be stationary, since the dictionary atoms capture the variation in the spectral characteristics. 

\subsection{Contributions of this work}
The main contributions and the novelty of the paper are:
\begin{itemize}
\item 
Using distinct dictionaries with each dictionary representing an audio source as well as a concatenated dictionary.

\item 
Dictionary learning by using thresholds on the cosine similarity to ensure  distinction amongst the atoms of the same as well as different source dictionaries.
\item 
Proposing two new objective measures, namely, the number of non-zero weights and the sum of weights, for selecting the most likely audio source from a  given set.
\end{itemize} 

\section{Proposed method}

\subsection{Problem formulation}

Given a test audio signal $s[n]$, we need to identify the signal as belonging to one of the noise or instrument sources. We train $M$ dictionaries for the $M$ different sources and the test audio signal is classified as that source which has the highest value for an objective measure. 
 
%
%

\subsection{Dictionary learning}
Similar to most of the audio source separation approaches \cite{virtanen0,ozerov,mysore}, the magnitude of short-time Fourier transform (mag. STFT) has been used as the feature vector, which is always non-negative. Feature vectors are $L2$ normalized for dictionary learning. A test feature vector can be represented as additive, non-negative, linear combination of dictionary  atoms.

	Each dictionary atom  is selected to be as uncorrelated as possible from the rest of the atoms belonging to the same as well as other sources. The correlation between a pair of atoms $\mathbf{d_n},\mathbf{d_j}$ is measured using the  cosine similarity   as:

\begin{equation}
cs(\mathbf{d_n},\mathbf{d_j})=\mathbf{d_n}^T\mathbf{d_j}/(||\mathbf{d_n}||||\mathbf{d_j}||)
\end{equation}
Two types of cosine similarity measures are used: (a) intra-class cosine similarity (intra-CS) is defined as  $cs_i(\mathbf{d_n},\mathbf{d_j}),\; \mathbf{d_n},\mathbf{d_j}\in D_k , n\neq j$
where $D_k$ is the dictionary for a specific source; and (b) inter-class cosine similarity (inter-CS) defined as $cs_I(\mathbf{d_n},\mathbf{d_j}),\; \mathbf{d_n}\in D_k,\,\mathbf{d_j}\in D_m, k\neq m $.

Dictionary atoms for each source are learnt such that the cosine similarity between the atoms is below a set threshold, chosen based on the desired performance. A randomly selected  feature vector, denoted as $\mathbf{f_r}$ is taken as the first atom for the first source, $\mathbf{d_1^1}$ . The rest of the atoms are learnt by  random selection of the feature vectors (excluding features already selected as atoms): $t^{th}$ feature, $\mathbf{f_t}$, is selected as the $n^{th}$ atom, $\mathbf{d^1_n}$ of dictionary $D_1$ if maximum of intra-CS, $max \;cs_{i}(\mathbf{f_t},\mathbf{d_j^1}),\; j< n$ (similar to coherence in \cite{tropp}) is less than a threshold $T_i$.

The selection of dictionary atoms is stopped once the number of dictionary atoms reaches a pre-determined number $N_{A}$. In case $N_{A}$ atoms are not obtained, additional mag. STFT features, which do not satisfy the intra-class threshold $T_{i}$ are appended in the order of increasing $max \;cs_i$.

For learning dictionaries for subsequent sources, atoms are learnt using an additional constraint: $\mathbf{f_t}$ is selected as the $n^{th}$ atom $\mathbf{d_n^k}$ for the $k^{th} $ dictionary $D_k$, if  $max\; cs_I(\mathbf{f_t}, \mathbf{d_j^h}),\; \mathbf{d_j^h}\in D_h,\:h<k, 1\leq j \leq N_A $  is less than a threshold $T_{I}$.

The threshold  $T_i$  ensures that the atoms within the same source dictionary are as uncorrelated as possible, while $T_I$ ensures that atoms from different source dictionaries are maximally uncorrelated. Lower the values of the thresholds $T_i$ and $T_I$, greater is the uncorrelatedness amongst the dictionary atoms. 

\begin{figure}
\centering 

\includegraphics[width=.5\textwidth,height=.2\textheight]{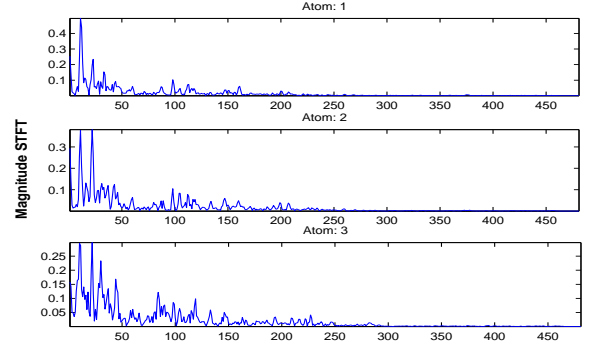}

\caption{The first three atoms from the dictionary of babble noise, learnt using  $T_i= 0.95,\; T_I=0.95$ }
\label{atomsnois}
\end{figure}

  The proposed source classification method has been evaluated using ten different noise sources taken from Noisex database \cite{noisex} and two other instrument sources, one recorded by us and  the other, downloaded from an open source portal \cite{veena}. The total number of atoms in $D$ from the 12 sources is 1200 using $T_i=T_I=0.95$ and $N_{A}=100$. For the sake of illustration, Fig. \ref{atomsnois} shows the plots of the first three atoms of babble noise learnt  for $T_i = T_I = 0.95$. The proposed DL is summarized in Algorithm \ref{dictalgo}. For the sake of simplicity, the algorithm does not show the appending of additional dictionary atoms when $N_{A}$ atoms could not be obtained.
\\
\begin{algorithm}{Dictionary learning}\label{dictalgo}
\begin{algorithmic}[1]
\State  \textbf{Initialize:}	Dictionary index $k=1$;  $D_k=\mathbf{d_1^1}=\mathbf{f_r}$;  Atom index $n=2$; set $T_i$ and $T_I$.
	
	\Repeat
	\State Extract $N$ number of  mag.STFT features denoted as $\mathbf{f_l}, 1\leq l\leq N$ from the $k^{th}$ audio source.
	\Repeat
	
	\State If $n>1$, find the maximum of intra-CS, $m_i$ as:
	
	 $\max( cs_i(\mathbf{f_t},\mathbf{d^k_j})\;\forall\: j=1...n-1)$
	 	 
\State	If $k>1$, find the maximum of inter-CS, $ m_I$   as:
		
		 $\max(cs_I(\mathbf{f_t},\mathbf{d^h_j})\:\forall\: j=1..N_A,h< k)$
 \If{$m_i\leq T_i$  and  $m_I\leq T_I$ (for $k>1$)}
\State Assign randomly selected $\mathbf{f_t}$ as the $n^{th}$ atom: $\mathbf{d^k_n}=\mathbf{f_t}$ and append to the dictionary: $D_k=[D_k\; \mathbf{d^k_n}]$
\State $n=n+1$

\EndIf

\Until{$n>N_{A}$ or  all $f_t$ are selected}
		
	\State $k=k+1$; $n=1$
\Until{All source dictionaries are learnt}
\end{algorithmic}
\end{algorithm}

\subsection{Classification stage}
 
The learnt dictionaries are used to extract measures for identifying the source. Given an unknown audio signal, the mag. STFT features are extracted, which are used to solve a minimization using ASNA \cite{virtanen}:
 
 \begin{equation}
\stackunder{minimize}{\textbf{x}}\;
 KL(\mathbf{y}||\mathbf{\hat{y}}), \; \mathbf{\hat{y}}=D\mathbf{x } \;s.t.\;x\geq 0 \label{asna}
 \end{equation}
where $KL()$ is the KL divergence between two vectors, $y$ is the extracted feature, $\mathbf{\hat{y}}$ is the approximation of $\mathbf{y}$, $ D$ is the dictionary using which $\mathbf{y}$ is approximated and $\mathbf{x}$ is the weight vector estimated using ASNA.
  
\begin{figure}
     \centering
    
     \includegraphics[width=.5\textwidth,height=.21\textheight]{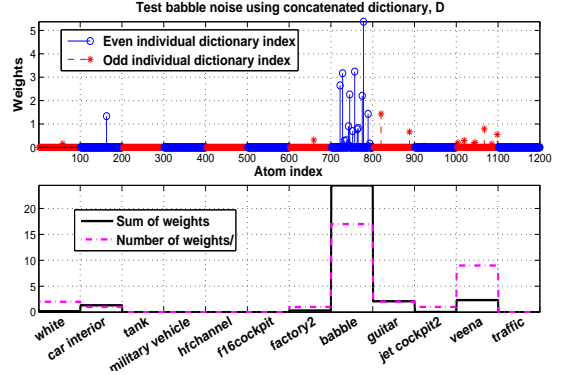}
     
     \caption{(a) Weights for a single frame of babble noise estimated by ASNA using concatenated dictionary, $D$. (b) Number and  sum of non-zero weights in (a) as a function of dictionary type for $T_i= T_I=0.95$.}
     \label{weignosum}
     \end{figure}

Since we know the dictionaries for all the sources, we estimate three measures for classification:
  
  \begin{enumerate} [leftmargin=*]
\item   \textit{Signal to distortion ratio} (SDR) \cite{vincent} between $\mathbf{y}$ and $\mathbf{\hat{y_i}}=D_i\mathbf{x_i},\;1\leq i\leq M$ for the $M$ dictionaries. The SDR with respect to each dictionary $D_i$ is defined as :
  \begin{equation}
  SDR_i=20\times log_{10}(||\mathbf{y}||_2/||\mathbf{y}-\mathbf{\hat{y_i}}||_2)  
  \end{equation}

A feature $\mathbf{y}$ belonging to the $k^{th}$ source can be approximated to a good accuracy by atoms belonging to $D_k$, since $D_k$ has been learnt by threshold based selection of atoms from the same source. So, $||\mathbf{y}-\mathbf{\hat{y}}||_2$ is expected to  be minimum for the $k^{th}$ source, since $\mathbf{y}$ may not be approximated well by atoms from the dictionaries of other sources. Thus, the $SDR_i$ is expected to be maximum for the $k^{th}$ dictionary. The estimated source index $k$ for the feature vector of each frame of the test signal is given as
$k=\stackunder{arg max}\;\;  SDR_i$. 
  
  \item  We propose a new measure, \textit{Number of non-zero weights} (NNZ) belonging to a particular source in the weight vector $\mathbf{x}$ recovered using a dictionary $ D$, obtained by concatenating dictionaries from all the $M$ individual sources: $ D= [D_1\: D_2 ... D_{M} ]$. The vector $\mathbf{x}=[\mathbf{x_1}'\:\mathbf{x_2}' ... \mathbf{x_M}']'$ obtained by using ASNA on (\ref{asna}) is a concatenation of individual weight vectors $\mathbf{x_i}$ of $M$ sources, which is expected to be sparse.

A test feature $\mathbf{y}$ belonging to the $k^{th}$ source can be  represented better by atoms from the  $k^{th}$ dictionary than by atoms from other dictionaries. Since $D$ contains atoms from all the sources, the number of non-zero weights, $NNZ_k$ corresponding to the original dictionary $D_k$, which is now a sub-matrix of $D$, may be expected to be higher than $NNZ_i,  i\neq k$. The estimated source index $\hat{k}$ for the test feature $\mathbf{y}$ is given by 
   $\hat{k}=\stackunder{arg max}\;\;  NNZ_i,\;1\leq i\leq M$.  
   
   The weight vector $\mathbf{x}$ is sparse for the dictionary $D$, as shown in Fig. \ref{weignosum}(a). The number of non-zero weights for each source dictionary is illustrated in Fig. \ref{weignosum}(b). For a test frame of babble noise, the highest NNZ is 17 corresponding to babble noise dictionary (atom indices 700 to 800 in $D$), while 9 is the next highest for the veena dictionary, a margin of 8 or a factor of 2, for correct classification is obtained.
   
   \item \textit{Sum of weights} (SW) is another scalar measure proposed, defined as the sum of the elements of the vector $\mathbf{x_{i}}$, which is recovered using the same concatenated dictionary, $ D$. In case the weights are non-sparse, it is observed that $SW_i$ is more reliable than $NNZ_i$. Figure \ref{weignosum}(b)  also illustrates the distribution of SW for each of the dictionaries. $\hat{k}=\stackunder{arg max}\;\;  SW_i$ gives the estimated source index for a test feature $\mathbf{y}$. The sum of weights is the highest (24.47) for babble noise dictionary, while that of veena is 2.33, a factor of about 10.5 for correct classification. It is to be noted that the dictionary used for both $NNZ$ and $SW$ is a concatenated dictionary $D$, while the measure $SDR$ is derived using separate dictionaries $D_i$. 
  \end{enumerate} 
    
  \section{Results and discussion}
 
Magnitude STFT features are extracted using a frame size of 60 ms and a frame shift of 15 ms from each audio source with a duration of 3 to 4 minutes. We experimented with different choices and arrived at these values as the optimum. Since the number of atoms in each dictionary is constrained to be 100, only 6 seconds from the training set of each audio type form the dictionaries. For evaluating the method, a test signal of duration 5 seconds, equivalent to 330 frames, is taken from the database, and the rest of the audio signal is used in the training stage for learning the dictionaries. 
         
\begin{figure}[!h]
      \centering
     
      \includegraphics[width=.5\textwidth,height=.18\textheight]{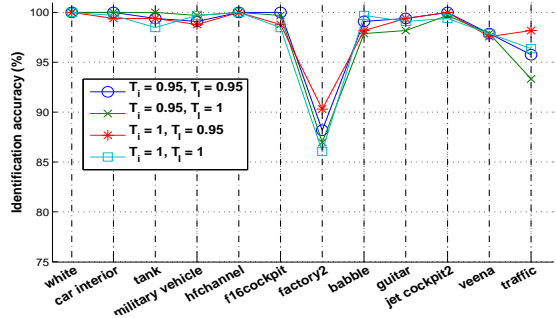}
      
      \caption{Percentage of 60 ms frames correctly detected as the original audio source using SDR as a measure, for different choices of $T_i$ and $T_I$.}
      \label{percentfrms}
      \end{figure}

\begin{figure}[!h]
        
        \hspace{-.4cm}     \includegraphics[width=.5\textwidth,height=.15\textheight]{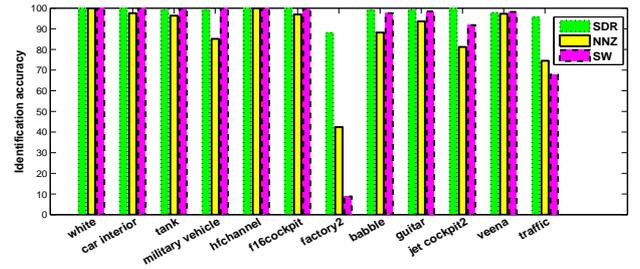}
             
             \caption{Individual classification accuracies for all the sources using the three measures independently.}
             \label{accallnois}
        \end{figure}

  Figure \ref{percentfrms} shows the plot of percentage of frames of each test signal correctly classified using SDR as the classification measure for various combinations of $T_i$ and $T_I$. Table \ref{overiden} summarizes the overall audio classification accuracy for different choices of $T_i$ and $T_I$, where the highest accuracy is obtained for $T_I=T_i=0.95$ using any of the measures SDR, NNZ and SW. Random selection of mag. STFT features along with the constraint on the cosine similarity has ensured distinct dictionaries and the capture of the variations in the audio characteristics by the atoms. The misclassification is marginally higher when either of the thresholds is unity. So, we have used $T_i=T_I=0.95$ as the thresholds. Figure \ref{accallnois} shows the percentage of frames correctly classified from each of the 12 audio sources for each of the three measures.
                            \begin{table} 
                   \centering   \caption{Overall source classification accuracy (\%) for different choices of $T_i$ and $T_I$ using SDR, NNZ and SW as measures }                                                                                          \begin{tabular}{|cc||c|c|c|}
                         \hline                                                                                        $T_i$ & $T_I$ & SDR & NNZ & SW \\
                                                                                                                \hline
                                                                                                                0.95 & 0.95 & 98.23 & 87.78 & 88.51 \\
                                                                                                                \hline
                                                                                                                0.95 & 1.00 & 98.01 & 87.13 & 88.01 \\
                                                                                                                \hline
                                                                                                                1.00 & 0.95 & 98.11 & 87.05 & 88.21 \\
                                                                                                                \hline
                                                                                                                1.00 & 1.00 & 98.06 & 87.03 & 88.42 \\
                \hline                                                                                                \end{tabular}
            \label{overiden}                                                                                                                                                                                                \end{table}   
  Even though SDR outperforms the other two measures, NNZ and SW are promising since they are computationally simple and give a different insight into the distribution of weights. In case  the number of audio sources $M$ is large, using only SDR as an classification measure is computationally complex, since ASNA is run $M$ number of times. In that case, the measures $NNZ$ or $SW$ can act as the front end for classification (since ASNA is run only once). These measures can pick up the top few source dictionaries and then, SDR can be used to find the best fit among them.

   Two higher level measures are defined for the $i^{th}$ dictionary, namely, accumulated SDR (ASDR) and moving ASDR (MASDR) as:
  
  \begin{equation}
ASDR_i(q)= \sum_{j=1}^{q}SDR_i(j)
  \end{equation}
  \begin{equation}
MASDR_i(q)= \sum_{j=q-P+1}^{q} SDR_i(j)
  \end{equation}
  
where $q$ is the index of the present frame and $P$ is the number of frames accumulated.
  
\begin{figure}
           \centering
        
          \subfigure[Test frames from factory noise]{
             \includegraphics[width=.5\textwidth,height=.14\textheight]{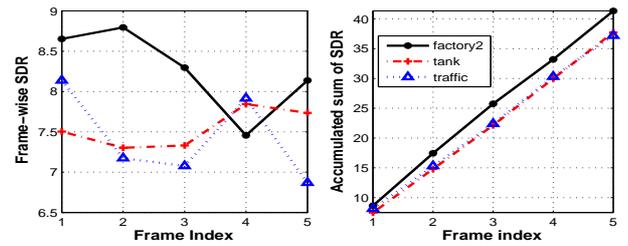}
         }
        
           \subfigure[Test frames from traffic noise]{
             \includegraphics[width=.5\textwidth,height=.14\textheight]{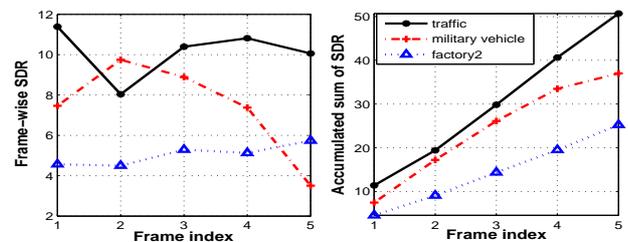}}
             
                \caption{Accumulated and frame-wise SDR for $T_i=T_I=0.95$ for test frames of factory and traffic noise.}
                \label{accsum}
                \end{figure}

  Figure \ref{accsum} shows the  frame-wise SDR and the corresponding ASDR  for five test frames of factory and traffic noise (most misclassified audio sources in Fig.\ref{percentfrms}). In each case, only two other audio sources having highest SDR's are shown, for clarity. It is seen in Fig.\ref{accsum}(a) that even though frame-wise SDR for the fourth frame is lower for factory noise, the corresponding ASDR is higher and gives correct classification.  In our experiment, we find that 100\% classification accuracy can be obtained using MASDR with $P=6$ for ten of the sources implying that  any consecutive six frames (135 ms) of the test noise are sufficient for correct classification. Test factory noise requires $P=10$ and veena, $P=14$ for correct classification.

   In a real life scenario, the accuracy of classification based on accumulated classification measures is more relevant than individual frame level accuracy, since the classification algorithm gets a stream of test audio signal as input. So,  even though a few frames may be individually misclassified, the accumulated  classification measure correctly classifies the source.

\subsection{Comparison with previous work}
                  Maleh et. al \cite{maleh} performed frame-wise noise identification (frame size of 20 ms) using line spectral  frequencies as features and pattern recognition based classifiers. They trained using 18.75 minutes of audio data each from 5 noise classes (three of them from NOISEX database), and tested on 500 frames of data for each class. Chu et. al \cite{chu} obtained an overall accuracy of 83.9\% in recognizing 14 environmental sounds. We have used 12 classes, and obtained an overall frame level accuracy of 98.2\% using SDR,  compared to 89\% reported in  \cite{maleh}. The highest accuracy given by majority vote classifier in \cite{giannoulis} is around 78\%.   The accuracy is 100\% using MASDR. 
  \section{Conclusion and future work}
  A new approach to audio source classification has been proposed adopting ASNA as the source recovery algorithm. Experiments using very limited  training data have shown a good overall frame level accuracy of  98\%. We plan to explore and devise other source recovery algorithms for faster and
  more efficient background source classification. 
  Also, we are working on classification of type of background noise from noisy speech and the subsequent separation of speech.

  \eightpt
  \bibliographystyle{IEEEtran}



\end{document}